\documentclass[aps,pra,twocolumn,groupedaddress,showpacs,superscriptaddress]{revtex4-1}
\usepackage{graphicx,amsmath}
\usepackage{amssymb}

\begin{document}
\title{Accelerated Quantum-Assisted Selected Configuration Interaction via Fast-Annealing-Based Determinant Selection}
\author{Hayun Park}
\affiliation{Department of Liberal Studies, Kangwon National University, Samcheok, 25913, Republic of Korea}
\author{Younghun Kwon}
\affiliation{Department of Applied Physics, Hanyang University, Ansan 19588, Republic of Korea}
\author{Hunpyo Lee}
\email{*Email: hplee@kangwon.ac.kr}
\affiliation{Department of Liberal Studies, Kangwon National University, Samcheok, 25913, Republic of Korea}
\affiliation{Quantum Sub Inc., Samcheok 25913, Republic of Korea}
\date{\today}

\begin{abstract}
Full configuration interaction (FCI) provides the exact electronic structure within a given atomic basis, but its 
computational cost grows exponentially with the number of spin orbitals. Selected configuration interaction (SCI) 
methods alleviate this limitation by retaining only the most important Slater determinants; however, the repeated 
identification of important determinants remains a major computational bottleneck. Here, we present 
a quantum-assisted selected configuration interaction (QA-SCI) method that combines SCI with graph-based block 
diagonalization (GBBD) of the FCI Hamiltonian. The GBBD method partitions the FCI Hamiltonian into independent 
blocks, within which the determinant-selection problem is formulated as a quadratic unconstrained binary 
optimization (QUBO) problem. The QUBO problems for selecting determinants to construct the SCI space are 
iteratively solved using a fast-annealing approach. The eigenvalues of the resulting QA-SCI Hamiltonian are 
computed using the Lanczos algorithm. We benchmark the proposed method on H$_8$–H$_{18}$ hydrogen chains and 
Li$_2$S in the STO-3G basis. For the H$_n$ chains, chemical accuracy is achieved while retaining only a small 
fraction of the Slater determinants, and this fraction systematically decreases with increasing chain length, 
despite the exponential growth of the FCI Hilbert space. For Li$_2$S, the QA-SCI results remain within chemical 
accuracy while retaining substantially fewer determinants than the full FCI space. Finally, we apply QA-SCI to 
N$_2$ using the 6-31G basis, considering both active-orbital and full-orbital treatments. The full-orbital QA-SCI 
calculation, using only approximately 50,000 Slater determinants, yields a lower ground-state energy than an FCI 
calculation within an active space comprising 12 spin orbitals and 12 electrons. These results demonstrate that the 
combination of QA-SCI and the GBBD approach can substantially reduce the computational cost of determinant 
selection while maintaining the accuracy of FCI-based electronic structure calculations, thereby enabling accurate 
calculations in larger orbital spaces.
\end{abstract}

\maketitle

\section{Introduction}

Reliable predictions of ground- and excited-state energies of molecules are essential for a wide range of 
applications, including catalyst design, drug discovery, and the development of novel materials. Among the 
available electronic structure methods, full configuration interaction (FCI) provides the exact solution to the 
electronic Schrödinger equation within a given one-electron basis set by considering all possible electronic 
configurations and fully accounting for electron correlation~\cite{knowles1984,szabo1996}. However, the dimension 
of the Hilbert space grows exponentially with the number of spin orbitals, making FCI calculations computationally 
prohibitive for all but relatively small molecular systems.

Two major approaches have been developed to extend FCI calculations to larger molecular systems while retaining the 
advantages of the FCI Hamiltonian. The first approach partitions the FCI Hamiltonian into smaller blocks without 
introducing additional approximations, allowing the ground- and excited-state energies to be obtained through the 
independent diagonalization of each block. Although this approach yields results equivalent to conventional FCI 
while requiring significantly less memory, its applicability remains limited by the number of spin orbitals because 
the dimension of each block still grows exponentially with system size~\cite{park2025jcp}. The second approach is 
selected configuration interaction (SCI), which retains only the most important Slater determinants, thereby 
substantially reducing the computational cost and memory requirements while preserving much of the accuracy of FCI. 
In SCI methods, important Slater determinants are identified using various strategies, including perturbative 
selection, adaptive ranking based on approximate wavefunction coefficients, heat-bath criteria, and stochastic 
sampling approaches such as Monte Carlo Configuration Interaction~\cite{huron1973,garniron2018,garniron2019,
ohtsuka2017,holmes2016,sharma2017,holmes2017,li2018,tubman2020,zhang2020,yao2020,damour2021}. 
Although these methods significantly reduce the size of the configuration space compared with FCI, they still 
require repeated identification of important Slater determinants and iterative 
diagonalization of the Hamiltonian~\cite{stair2020,eriksen2021,prentice2023}. Furthermore, the computational cost 
associated with Slater determinant 
selection increases rapidly with the number of spin orbitals and eventually becomes a major computational 
bottleneck. Therefore, developing efficient SCI approaches that can accurately and fast identify important Slater 
determinants within each Hamiltonian block remains highly desirable.

Recently, various quantum computing platforms have been developed for solving computationally demanding 
optimizationproblems and have also been employed as quantum eigensolvers and samplers for molecular electronic 
structurecalculations~\cite{abrams1999,aspuru2005,peruzzo2014,kanno2023,nakagawa2024,mikkelsen2025,
shirai2025,weaving2026,robledo2025,pellow2025}. Among these platforms, D-Wave 
quantum annealers are particularly well suited for solving quadraticunconstrained binary optimization (QUBO) 
problems through quantum annealing with short annealing
times~\cite{kadowaki1998,johnson2011,teplukhin2020,teplukhin2021}. Unlike gate-based quantum computers, which rely 
on sequences of quantum gate operations, quantum annealers directly search for low-energy solutions to optimization 
problems by exploiting quantum annealing dynamics. We guess that since determinant selection is a major 
computational bottleneck in SCI methods, reformulating this task as a QUBO problem can enable quantum annealing to 
accelerate the determinant-selection process.

In this work, we propose an accelerated quantum-assisted selected configuration interaction (QA-SCI) approach based 
on the graph-based block diagonalization (GBBD) method for FCI Hamiltonians constructed using the OpenFermion 
package, which was recently developed by two of the authors, in combination with quantum optimization. In the GBBD 
method, Slater determinants and nonzero off-diagonal Hamiltonian matrix elements are represented as the nodes and 
edges of a graph, respectively~\cite{park2025jcp}. Independent Hamiltonian blocks are then efficiently identified 
as connected components of the graph representation of the FCI Hamiltonian. Following the GBBD procedure, the 
selection of important Slater determinants within each Hamiltonian block is 
reformulated as a QUBO problem using a randomly sampled subset of Slater determinants. The resulting QUBO problem 
is optimized to identify important Slater determinants, and the selected determinants are iteratively expanded to 
efficiently identify dominant electronic configurations. Consequently, the Slater determinant-selection procedure 
can be partially offloaded from the classical processor to the D-Wave quantum annealer, reducing the computational 
cost of determinant selection while maintaining the accuracy of the SCI calculation. Finally, the QA-SCI 
Hamiltonian is constructed from the selected Slater determinants, and its eigenvalues are obtained using the 
Lanczos diagonalization algorithm.

To evaluate the accuracy and efficiency of the proposed QA-SCI method, we first consider hydrogen chains (H$_8$–
H$_{18}$) in the STO-3G basis as benchmark systems, as their electronic structures are relatively simple and well 
understood. Specifically, the electronic structures of H$_{16}$ and H$_{18}$ cannot be computed using FCI on our 
computational resources due to memory limitations. We therefore analyze the number of Slater determinants required 
to achieve chemical accuracy for the H$_n$ chains up to $n=14$, for which exact FCI results are available. In 
addition, the ground-state energies and energy gaps between the ground and first excited states are investigated as 
functions of the chain length $n$. The results for H$_{16}$ and H$_{18}$ are obtained using the QA-SCI approach and 
compared with the trends established by the FCI results for chain lengths up to $n=14$. We find that the QA-SCI 
results exhibit physically reasonable trends. Next, we further demonstrate the applicability of the proposed QA-SCI 
method by performing an electronic structure calculation for the Li$_2$S molecule in the STO-3G basis and comparing 
the results obtained using FCI and SCI. Finally, we consider N$_2$ using the 6-31G basis with both 
an active-orbital (AO) treatment and a full-orbital (FO) plus SCI treatment. We find that the ground-state energy 
obtained with FO plus SCI is lower than that obtained using an AO treatment comprising 12 spin 
orbitals and 12 electrons. These results demonstrate that the QA-SCI method combined with the GBBD approach can 
substantially reduce the computational cost of determinant selection through quantum annealing while retaining the 
accuracy of FCI-based electronic structure calculations.

\section{Method}
The starting point of the proposed QA-SCI approach is the construction of the FCI Hamiltonian in the 
atomic orbital basis using the second-quantized representation. The FCI Hamiltonian is given as
\begin{equation}
\hat{H} = E_{\mathrm{nuc}} + \sum_{pq} h_{pq}, a_p^{\dagger} a_q + \frac{1}{2} \sum_{pqrs} h_{pqrs} a_p^{\dagger} 
a_q^{\dagger} a_s a_r,
\label{eq:hamiltonian}
\end{equation}
where $E_{\mathrm{nuc}}$ is the nuclear repulsion energy, $a_p^\dagger$ and $a_p$ are the fermionic creation and 
annihilation operators acting on spin orbital $p$, and $h_{pq}$ and $h_{pqrs}$ denote the one- and two-electron 
integrals, respectively. Here, these integrals are generated using the OpenFermion package~\cite{mcclean2020}. 
Because the dimension of the FCI Hamiltonian grows exponentially with the number of spin orbitals, both the 
computational cost and memory requirements increase rapidly. To alleviate these computational 
challenges, we first employ the recently developed GBBD method to partition the 
FCI Hamiltonian of Eq.~(\ref{eq:hamiltonian}) into multiple independent subblocks~\cite{park2025jcp}. 
The ground- and excited-state eigenvalues are then computed independently for each subblock, substantially reducing 
the computational cost compared with the exact diagonalization of the full FCI Hamiltonian.

To further reduce the computational cost within each subblock, we introduce the QA-SCI method, 
in which the process of important Slater determinants is formulated as a QUBO problem. The corresponding QUBO 
objective function is defined as
\begin{equation}
Q(\mathbf{x}) = \mathbf{x}^{\mathrm T}\hat{H}\mathbf{x}+\lambda\left(\sum_i x_i-N\right)^2,
\label{eq:qubo_base}
\end{equation}
where $\mathbf{x}$ is a binary vector with components $x_i\in{0,1}$. The penalty parameter $\lambda$ enforces the 
cardinality constraint, and $N$ denotes the target number of selected Slater determinants. Each binary variable 
$x_i$ indicates whether the $i$-th Slater determinant is retained $(x_i=1)$ or discarded $(x_i=0)$ in the SCI 
calculation. The optimal subset of Slater determinants is obtained by solving
\begin{equation}
\min_{\mathbf{x}\in{0,1}} Q(\mathbf{x}).
\label{eq:qubomin}
\end{equation}
The QUBO formulation is designed to be compatible with D-Wave quantum annealers, which can perform quantum 
annealing with annealing times as short as approximately $100~\mu\mathrm{s}$. Although all numerical calculations 
presented in this work are performed using simulated annealing with the D-Wave Ocean package, the same QUBO 
formulation can be directly implemented on D-Wave quantum annealers. Such rapid quantum annealing has the potential 
to accelerate the Slater determinant selection process and reduce the computational cost of SCI calculations. We 
refer to the proposed method as quantum-assisted because the determinant-selection problem is formulated as a QUBO 
that is directly compatible with quantum annealing hardware, while the numerical results presented in this work are 
obtained using simulated annealing. Because implementing a fully connected QUBO on D-Wave quantum annealers 
requires embedding onto the hardware connectivity graph, we restrict the QUBO problems to fully connected instances 
with approximately 200 logical variables~\cite{park2024avs}. We randomly sample subsets of 200 binary variables 
from $\mathbf{x}$ and repeatedly solve the corresponding QUBO problems to identify and accumulate important Slater 
determinants~\cite{park2024prb,park2025npsm}. Finally, a reduced QA-SCI Hamiltonian is constructed from the Slater 
determinants selected through repeated QUBO optimization, and its eigenvalues are computed using the Lanczos 
algorithm. The size of the reduced QA-SCI Hamiltonian is controlled by the number of QUBO optimization runs 
specified by the user. Increasing the number of optimization runs generally enlarges the reduced QA-SCI space by 
incorporating additional important Slater determinants, thereby systematically improving the accuracy of the SCI 
calculation. All computations were performed on a MacBook Pro equipped with an Apple M4 Pro processor (14-core CPU) 
and 48 GB of unified memory.

\section{Results}
\subsection{H$_n$ clusters}
In this subsection, we present benchmark results for H$_n$ in the STO-3G basis set. For an H$_n$ chain, the 
dimension of the full FCI Hilbert space is $2^{2n}$. By enforcing fixed numbers of $\alpha$- and $\beta$-spin 
electrons with symmetry, the Hamiltonian dimension is 
reduced to $\binom{n}{m_{\alpha}}\binom{n}{m_{\beta}}$, where $m=m_{\alpha}+m_{\beta}$.
This reduced Hamiltonian is first partitioned into independent blocks using the GBBD algorithm.
The proposed QA-SCI method is then applied independently to each block to identify the dominant
Slater determinants. There are two possible approaches for selecting important Slater determinants in SCI 
calculations. The first approach is to remove less important Slater determinants from each full FCI block 
Hamiltonian 
while retaining the dominant ones. The second one is to collect important Slater determinants by sampling randomly 
selected Slater determinants from each FCI block Hamiltonian. The results obtained using both approaches are in 
excellent agreement.

\begin{figure}
\includegraphics[width=0.95\columnwidth]{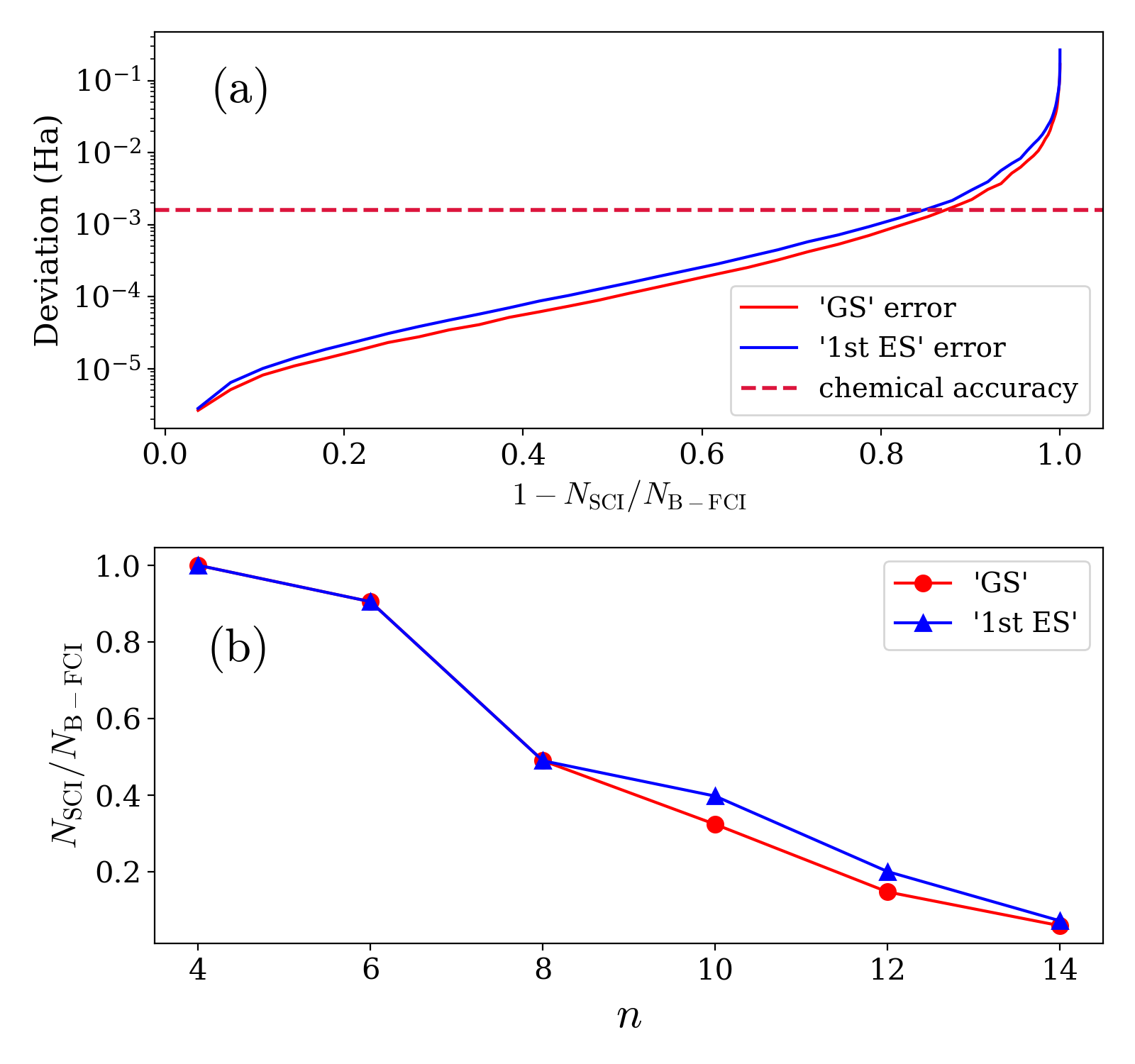}
\caption{\label{Fig1} (Color online) (a) Deviation of the ground- and first-excited-state energies from their 
corresponding full configuration interaction (FCI) values as a function of the fraction of removed Slater 
determinants, $1-N_{\mathrm{SCI}}/N_{\mathrm{B\mbox{-}FCI}}$, for the H$_{12}$ chain with a lattice constant of 
$0.9$~\AA. The Hamiltonian blocks corresponding to the ground and first excited states contain 427,008 and 426,688 
Slater determinants, respectively. (b) Fraction of Slater determinants removed while maintaining chemical accuracy, 
$N_{\mathrm{SCI}}/N_{\mathrm{B\mbox{-}FCI}}$, as a function of $n$, the number of H atoms, for the ground- and 
first-excited-state energies. Here, GS and 1st ES denote the ground state and first excited state, respectively.}
\end{figure}

Figure~\ref{Fig1}(a) presents the deviations of the ground- and first-excited-state energies from the 
corresponding FCI values as a function of the ratio $1-N_{\mathrm{SCI}}/N_{\mathrm{B\mbox{-}FCI}}$ for the 
H$_{12}$ chain with a lattice constant of $0.9$~\AA, where $N_{\mathrm{B\mbox{-}FCI}}$ denotes the dimension of 
the corresponding FCI Hamiltonian block. The numbers of FCI Hamiltonian blocks for the ground and 
first-excited states contain 427,008 and 426,688 Slater determinants, respectively. As expected, the deviations from the FCI 
energies increase monotonically as the fraction of retained Slater determinants decreases. Chemical accuracy is 
maintained when approximately 15\% of the Slater determinants are retained for the H$_{12}$ chain. The last 
remaining Slater determinant corresponds to the Hartree-Fock (HF) singlet configuration $\vert 111111111111000000000000 
\rangle$ for the ground state and the HF triplet configuration $\vert 111111111110010000000000 \rangle$ 
for the first excited state. Figure~\ref{Fig1}(b) shows the fraction of Slater determinants retained 
in each FCI block that is required to reproduce the ground- and first-excited-state energies of the H$_n$ clusters 
within chemical accuracy. Although the sizes of the FCI blocks increase exponentially with the number of $n$, 
following the same scaling as the dimension of the Hilbert space, the fraction of retained Slater determinants 
required to achieve chemical accuracy decreases systematically with increasing system size for both the ground- and 
first-excited state energies. This result indicates that an increasingly smaller fraction of the Hilbert space is 
sufficient to recover the energetics within chemical accuracy as the system size increases.

\begin{figure}
\includegraphics[width=0.95\columnwidth]{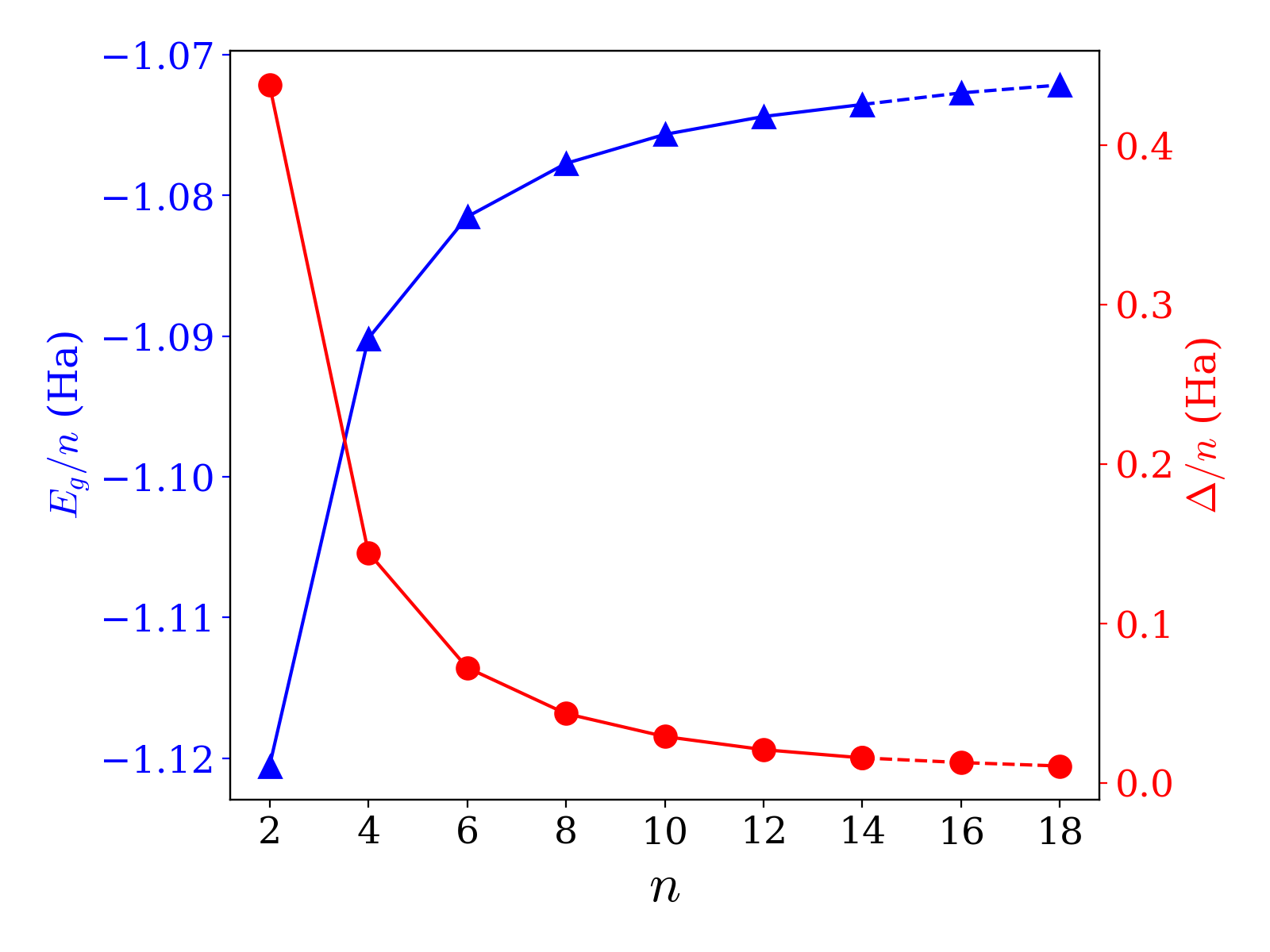}
\caption{\label{Fig2} (Color online) Ground-state energies $E_g/n$ and energy gaps $\Delta/n$ between the ground 
and first excited states as a function of the number of H atoms, $n$.}
\end{figure}

Figure~\ref{Fig2} shows the ground-state energies $E_g/n$ and energy gaps $\Delta/n$ between the ground 
and first excited states as a function of the number of H atoms, $n$. The dashed lines represent the exact FCI 
results for $n=2$ to $14$, while the dotted lines 
represent the QA-SCI results for $n=16$ and $18$. The exact FCI results for $n=16$ and $18$ are not available due 
to the memory limitations of our computational resources. Nevertheless, we expect the QA-SCI results for $n=16$ and 
$18$ to remain within chemical accuracy based on the trends observed for the smaller H$_n$ chains. The energy gap 
decreases toward zero as the chain length increases, indicating that the system approaches metallic behavior in the 
limit of an infinitely long H$_n$ chain.

\subsection{Li$_2$S and N${_2}$ molecules}
\begin{figure}
\includegraphics[width=0.95\columnwidth]{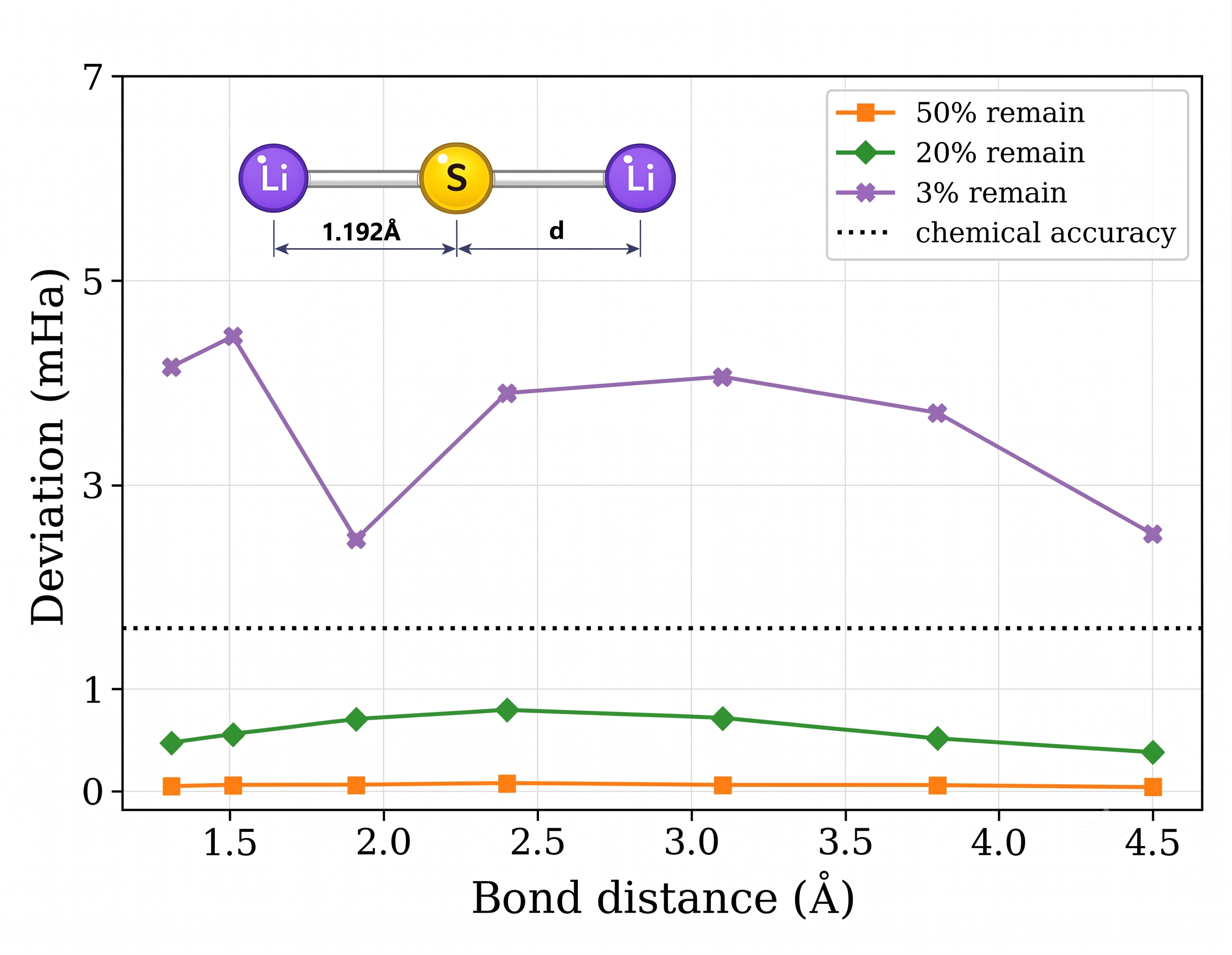}
\caption{\label{Fig3} (Color online) Deviation between the exact FCI and SCI results as a function of the bond 
distance $d$ in the STO-3G basis. Here, $d$ denotes the Li–S bond length while the position of the other Li atom is 
fixed. The dotted line indicates the threshold for chemical accuracy.}
\end{figure}

In this subsection, we further benchmark the QA-SCI method for the Li$_2$S molecule by comparing the FCI and SCI 
results for the ground-state energy within the ground-state block of the block-diagonalized Hamiltonian. 
Figure~\ref{Fig3} shows the deviation between the exact FCI and SCI results as a function of the bond distance $d$ 
in the STO-3G basis. Here, $d$ denotes the Li–S bond length while the position of the other Li atom is fixed. The 
dotted line indicates the threshold for chemical accuracy. When approximately 50\% of the 
total Slater determinants in the ground-state block are retained, the deviation is nearly zero. As the fraction of 
removed Slater determinants increases, the deviation also increases. Nevertheless, the deviation remains within 
chemical accuracy when approximately 20\% of the Slater determinants are retained.

\begin{figure}
\includegraphics[width=0.95\columnwidth]{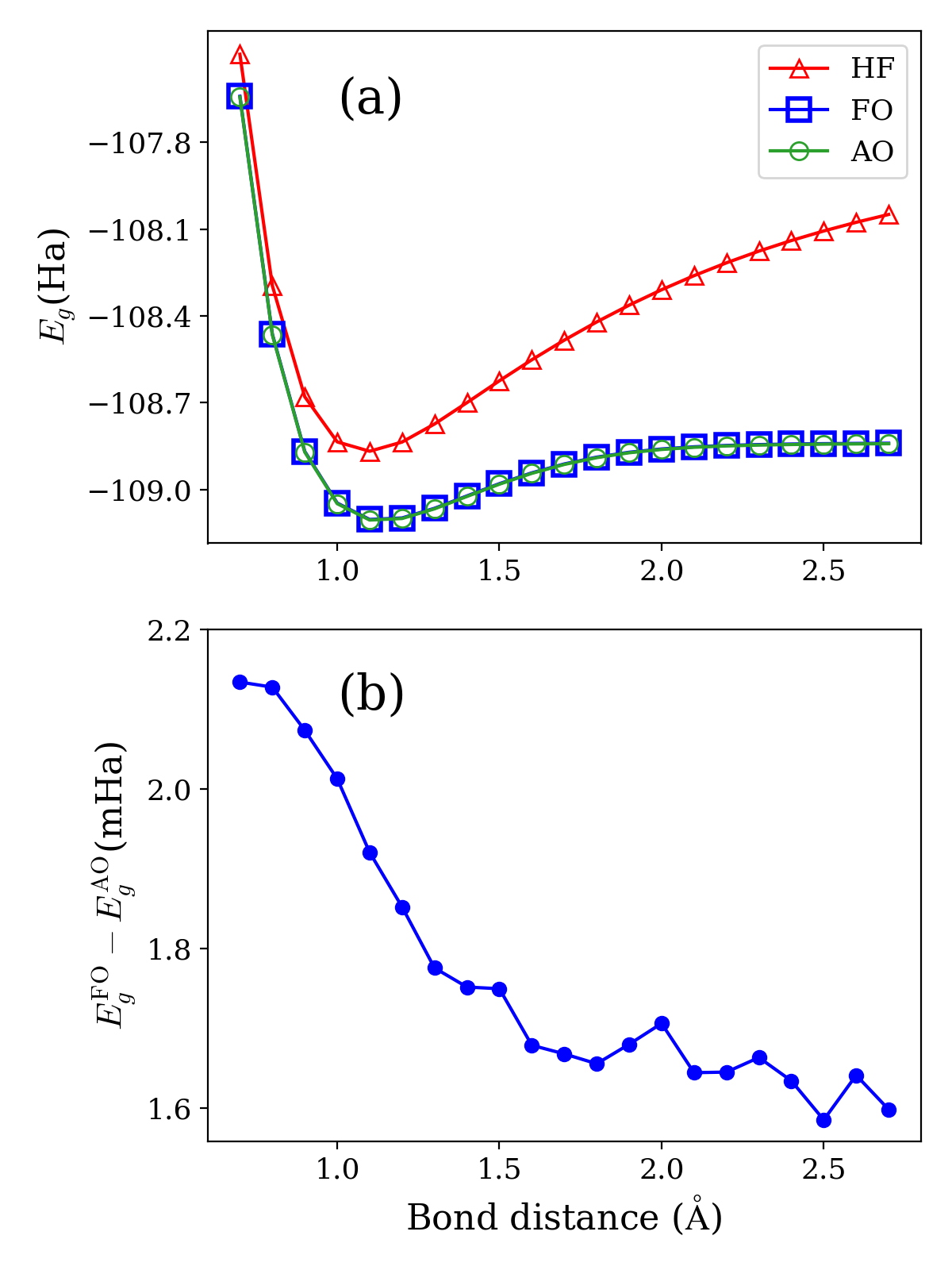}
\caption{\label{Fig4} (Color online) (a) Ground-state energies $E_g$ of N$_2$ calculated using the Hartree–Fock 
(HF) method with the 6-31G basis set, considering both active-orbital (AO) and full-orbital (FO) spaces. (b) Energy 
deviations between the AO and FO spaces obtained using the SCI approach as a function of bond distance, expressed 
on the mH scale.}
\end{figure}

Finally, we consider the N$_2$ molecule using the 6-31G basis to demonstrate the applicability of 
the proposed QA-SCI method to a more realistic system beyond the STO-3G basis. We consider both an AO space with 12 
spin orbitals and 12 electrons, for which the exact FCI results are available, and a FO space with 18 spin orbitals 
and 14 electrons, for which the exact FCI results are computationally 
inaccessible, to demonstrate the accuracy and efficiency of our approach. Figure~\ref{Fig4}(a) shows 
the ground-state energies $E_g$ of N$_2$ obtained using the HF method and 6-31G basis sets, including 
calculations with both AO and FO spaces. The HF energies are 
substantially higher than those obtained using the 6-31G basis, while the 6-31G results obtained with the different 
orbital spaces are nearly identical on the energy scale shown in Figure~\ref{Fig4}(a). To clearly resolve the 
energy differences between the 6-31G results, we plot the energy deviations between an AO space and 
FO space with SCI approach as a function of the bond distance on 
the mH scale in Figure~\ref{Fig4}(b). We find that the energies obtained using the SCI approach with the FO space 
are lower than those obtained using the FCI approach with the active-orbital space, despite the SCI calculation 
using only approximately 50,000 Slater determinants.

\section{Sumary}
We developed an accelerated QA-SCI approach combined with the GBBD method to efficiently compute the low-energy 
eigenvalues of the FCI Hamiltonian for molecular systems. The GBBD method not only reduces the computational burden 
but also separates the energy states into independent blocks, thereby preventing mixing between different states 
and facilitating the QA-SCI approach. A randomly sampled subset of Slater determinants from each Hamiltonian block 
is used to formulate a QUBO problem, which can be solved using quantum annealing. The important Slater determinants 
identified from the QUBO solutions are iteratively selected and accumulated to construct the SCI space. 
The ground- and low-lying excited-state energies are obtained using the Lanczos algorithm within the SCI space.

We examine the electronic structures of H chains ranging from H$_2$ to H$_{18}$. Specifically, FCI calculations 
for H$_{16}$ and H$_{18}$ are not feasible on our computational resources due to memory limitations. We find that 
although the size of the FCI space increases exponentially with the number of spin orbitals, the size of the SCI 
space grows much more slowly and does not follow the exponential scaling of the FCI space. In addition, to assess 
the accuracy of the proposed QA-SCI method, we compare the FCI results for H$_{14}$ with the QA-SCI results for 
H$_{16}$ and H$_{18}$ for both the ground-state energy and the energy gap. The results follow the trends 
established by the FCI calculations and are expected to remain within chemical accuracy. Furthermore, the energy 
gap decreases toward zero with increasing chain length, suggesting that it converges to zero in the limit of an 
infinitely long H$_n$ chain. We further assess the applicability of the proposed approach to the Li$_2$S molecule 
by comparing the energies obtained from FCI and SCI calculations. We find that the SCI results with 10,000 Slater 
determinants agree well with the FCI results within chemical accuracy. Finally, we compute the energies of the 
N$_2$ molecule using the 6-31G basis, considering both a limited AO space with FCI and the FO 
space with SCI. The AO space includes 12 spin orbitals and 12 electrons, whereas the FO space 
includes 18 spin orbitals and 14 electrons. We confirm that despite using a much smaller number of Slater 
determinants in the FO space, the SCI approach yields lower energies than the FCI calculation performed 
within the AO space.

\section{Acknowledgements\label{Ack}}
This work was supported by Institute of Information and communications Technology Planning Evaluation (IITP) grant
funded by the Korean government (MSIT) (No. RS-2023-0022952422282052750001) and by the ANCHOR program through the 
Gangwon ANCHOR center, funded by the Ministry of Education(MOE) and the Gangwon State(G.S), Republic of Korea 
(2026-ANCHOR-10-002).

\end{document}